\documentclass[doublecol]{epl2} 
\usepackage{amsmath}

\title{Kinetic Heterogeneities at Dynamical Crossovers}
\author{T. K. Haxton \and A. J. Liu}
\shortauthor{T. K. Haxton \etal}

\institute{Department of Physics and Astronomy, University of Pennsylvania, Philadelphia, PA 19104 USA}

\abstract{
We perform molecular dynamics simulations of a model glass-forming liquid to measure the size of kinetic heterogeneities, using a dynamic susceptibility $\chi_{\rm ss}(a, t)$ that quantifies the number of particles whose dynamics are correlated on the length scale $a$ and time scale $t$.  By measuring $\chi_{\rm ss}(a, t)$ as a function of both $a$ and $t$, we locate local maxima $\chi^\star$ at distances $a^\star$ and times $t^\star$.  Near the dynamical glass transition, we find two types of maxima, both correlated with crossovers in the dynamical behavior: a smaller maximum corresponding to the crossover from ballistic to sub-diffusive motion, and a larger maximum corresponding to the crossover from sub-diffusive to diffusive motion.  Our results indicate that kinetic heterogeneities are not necessarily signatures of an impending glass or jamming transition.
}

\begin{document}

\maketitle

\section{Introduction}
The dynamics in a liquid become increasingly correlated in space and heterogeneous in time as the system is cooled toward its glass transition.  These correlations are most pronounced at time scales on the order of the relaxation time of the system and are typically quantified using a dynamic susceptibility that measures fluctuations in the number of mobile particles~\cite{Glotzer2000b, Richert2002, Andersen2005, Berthier2005, Dalle-Ferrier2007, Berthier2007a, Berthier2007b}. A particle is deemed mobile if it has moved a distance of at least $a$ in a time interval $t$, so the dynamic susceptibility depends on the choice of $a$ and $t$.  For $a$ fixed at a fraction of the particle diameter, the dynamic susceptibility exhibits a maximum at times $t$ on the order of the relaxation time.  This maximum increases in height as temperature is lowered towards the glass transition and the relaxation time increases~\cite{Matsui1994, Kob1995, Matsui1998, Hiwatari1998, Glotzer2000, Lacevic2003b, Toninelli2005, Szamel2006, Chandler2006, Berthier2007a, Flenner2007, Charbonneau2007}.  Thus, the peak in the dynamic susceptibility is viewed as an important signature of the glass transition.

In molecular dynamics simulations of glass-forming liquids, the dynamic susceptibility is typically measured as a function of $t$ with the length scale $a$ fixed.  Studies of the length-scale dependence have been more rare~\cite{Lacevic2003b, Chandler2006, Abate2007, Charbonneau2007}.
However, recent experiments on granular systems~\cite{Lechenault2008} suggest that it is instructive to study the dynamic susceptibility as a function of $a$ as well as $t$.  This approach has the advantage of avoiding any arbitrariness in the choice of $a$ or $t$.

In this paper, we use the dynamic susceptibility to measure the spatial extent of kinetic heterogeneities as a function of distance $a$ and time $t$ in a model glass-forming liquid.  At temperatures where there is a well-defined sub-diffusive plateau in the root mean square displacement, we find that there are 
{\emph two} distinct local maxima in the dynamic susceptibility $\chi_{\rm ss}(a,t)$ quantifying the spatial extent of kinetic heterogeneities.  These two maxima correspond to crossovers in the dynamics.  The larger of the two maxima, which has been observed in previous simulations, occurs at late times at the crossover from sub-diffusive to diffusive motion.  This maximum quantifies the strong and well-studied correlations in the dynamics that arise at the scale of the relaxation time.  In addition, we find a secondary maximum at much earlier times, corresponding to the crossover from ballistic to sub-diffusive motion.  The presence of this second maximum indicates that the dynamics also exhibit significant spatial correlations at time scales corresponding to the trapping of a particle in the cage of its neighbors.  The discovery of a second maximum implies that kinetic heterogeneities are not just associated with the onset of slow dynamics near a glass or jamming transition, and that they should be regarded more generally as symptoms of crossovers in the dynamics.

\section{Model}
Our model consists of disks interacting in two dimensions via the purely repulsive pairwise potential
\begin{equation}\label{potential}
V(r_{ij})=
\begin{array}{lr}
\frac{\epsilon}{\alpha}\left(1-\frac{r_{ij}}{\sigma_i+\sigma_j}\right)^2 &\rm{for \ } r_{ij}<\sigma_i+\sigma_j,\\
0 &\rm{for \ } r_{ij}\ge \sigma_i+\sigma_j,
\end{array}
\end{equation}
where $\epsilon$ is the characteristic energy scale and $\sigma_i$ is the radius of the $i$th disk.  We use equal mixtures of disks of radii $\sigma$ and $1.4\sigma$ and equal mass $m$.  We choose units such that 2$\sigma$, $\epsilon$, m, and $k_B$ set equal to 1.  This sets the time unit $2\sigma\sqrt{m/\epsilon}=1$.  To solve the dynamics, we perform molecular dynamics simulations at fixed temperature and pressure in square periodic cells containing $N=100$, 400, or 1600 disks.  We employ Gaussian constraints\cite{Evans1983a} to fix the instantaneous temperature $T=|\vec{p}_i|^2/2(N-1)$ and hydrostatic pressure $p=(\vec{r}_i\cdot \vec{F}_i/2+NT)/L^2$, where $\vec{F}_i=-\vec\nabla\sum_jV(r_{ij})$, allowing the side length L of the periodic cell to vary.  In our runs, we fix $p=10^{-2}$ and  measure the dynamics at various temperatures $T=10^{-4}$ to $T=9\times 10^{-4}$ by running 10 to 20 simulations for durations between $\Delta\tau=10^5$ and $\Delta\tau=10^7$ after equilibrating at the temperature for one fifth as long.  We arrive at the temperature by quenching at rates $10^{-9}$ or $10^{-8}$ starting from well-equilibrated configurations at $T=10^{-3}$.  Results here are shown for a quench rate of $10^{-9}$.

\section{Overlap function and its correlations}
Correlations in the dynamics are measured by using the standard time- and distance-dependent order parameter 
\begin{equation}
q_{\rm s}(a, t; i, \tau)\equiv w_a(\vec{r}_i(\tau)-\vec{r}_i(\tau+t)),
\label{orderdef}
\end{equation}
where $w_a(\vec{r})$ is an overlap function~\cite{Glotzer2000}.  Here, $w_a(\vec{r})=1$ if $|\vec{r}|<a$ and $0$ otherwise.  We denote the average over all particles $i$ by $Q_{\rm s}(a, t; \tau)\equiv(1/N)\sum_{i} q_{\rm s}(a, t; i, \tau)$ and its time average by $\bar{Q}_{\rm{s}}(a, t)\equiv\langle Q_{\rm s}(a, t; \tau)\rangle_\tau.$  At very short times, $\bar{Q}_{\rm{s}}(a, t)=1$ because no particles have moved a distance $a$.   At very long times, positions at time $\tau+t$ are uncorrelated with positions at time $\tau$, so $\lim_{t\rightarrow \infty}\bar{Q}_{\rm{s}}(a, t)\simeq\phi a^2/N$, where $\phi$ is the packing fraction.  Between these two extremes, $\bar{Q}_{\rm{s}}(a, t)$ decays smoothly from 1 to $\phi a^2/N$.

The spatial extent of fluctuations can be characterized by the variance of the overlap function over different starting times~\cite{Glotzer2000}
\begin{equation}
\chi_{\rm ss}(a, t)\equiv N\left(\langle Q_{\rm s}(a, t; \tau)^2\rangle_\tau-\bar{Q}_{\rm{s}}(a, t)^2\right).
\label{chidef}
\end{equation}
As shown by the counting argument of ref.~\cite{Abate2007}, $\chi_{\rm ss}(a, t)$ is roughly the number of particles that move a distance $a$ over a time $t$ in a correlated manner.  

Note that the spatially averaged overlap function $Q_{\rm s}(a, t; \tau)$ is the self part of the order parameter $Q(a, t; \tau)\equiv(1/N)\sum_{ij} w_a(\vec{r}_i(\tau)-\vec{r}_j(\tau+t))$, and $\chi_{\rm ss}(a, t)$ is the self-self part of the dynamic susceptibility $\chi_4(a, t)\equiv N(\langle Q(a, t; \tau)^2\rangle_\tau-(\langle Q(a, t; \tau)\rangle_\tau)^2).$  Glotzer et al showed that the self-self part of the dynamic susceptibility dominates over the distinct-distinct and the self-distinct parts for a model glass-forming liquid~\cite{Glotzer2000, Lacevic2003b}.  

\begin{figure}[t]
\onefigure[width=80mm]{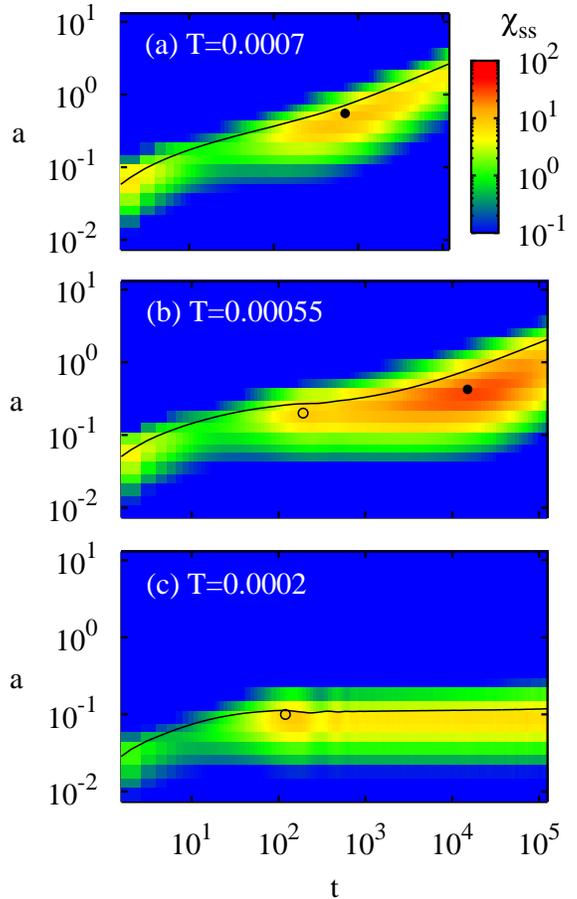}
\caption{Color plots of the value of the dynamic susceptibility $\chi_{\rm ss}(a, t)$ as a function of the lag time, $t$ and the overlap distance, $a$.  Results are shown at a fixed pressure, $p=10^{-2}$, and system size, $N=1600$, at three different temperatures: (a) $T=7\times 10^{-4}$, (b) $T=5.5\times 10^{-4}$, and (c) $T=2\times 10^{-4}$.  The dynamical glass transition is at $T \approx 5 \times 10^{-4}$.  For comparison, we plot the root-mean-square displacement $\Delta r(t)$ (solid curves).  The circles mark the locations, $(a^{\star}, t^{\star})$,  of observed local maxima of $\chi_{\rm ss}(a, t)$.  Filled circles represent cage-escaping local maxima near the end of the sub-diffusive plateau of $\Delta r(t)$.  Such maxima are observed for $T\ge5\times10^{-4}$.  Open circles represent cage-exploring local maxima near the beginning of the plateau.  Such maxima are observed for $T\le5.5\times10^{-4}$, so that there is a range of temperatures over which both local maxima exist.}
\label{maps}
\end{figure}

\section{Results}
The color plots of fig.~\ref{maps} display the spatial and temporal dependence of the dynamic susceptibility $\chi_{\rm ss}(a, t)$.  The three plots represent three temperatures:  one well above the dynamic glass transition $T_g\approx 5 \times 10^{-4}$, one slightly above it, and one below $T_g$.  In each plot, we also display the root-mean-square (rms) displacement,  $\Delta r (t)\equiv\sqrt{\langle|\vec{r}_i(0)-\vec{r}_i(t)|^2\rangle}$.  Not surprisingly, for a given $t$ we find that $\chi_{\rm ss}(a, t)$ is largest for $a$ near $\Delta r(t)$, the characteristic distance over which some, but not all, particles have moved. This is consistent with granular experiments by Lechenault, et al~\cite{Lechenault2008}.  We place circles on the locations $(t^\star, a^\star)$ of the local maxima of $\chi_{\textnormal ss}(a, t)$.  

For the highest temperature (fig.~\ref{maps}(a)), well above the glass transition, the rms displacement $\Delta r(t)$ shows a ballistic regime at small $t$, the hint of a sub-diffusive regime at intermediate times, and a diffusive regime at long times.   We observe one local maximum of $\chi_{\rm ss}(a, t)$, marked by a solid circle, at the onset of the diffusive regime: the heterogeneities are largest when some particles have begun to diffuse but others remain caged.  Note also that the maximum at $(t^*,a^*)$ lies somewhat below the $\Delta r (t)$ curve, indicating that larger clusters of dynamically correlated particles are formed by less mobile particles~\footnote{In all cases, we also find that $\chi_{\rm ss}(a, t)$ increases along $\Delta r(t)$ with decreasing $t$ for the shortest times observed, well within the ballistic regime, reflecting correlations in the velocities.  We observe this behavior at all temperatures.  This is an artifact of the barostat used; we do not observe this increase when the system is held at fixed area.}.  

For the intermediate temperature (fig.~\ref{maps}(b)), the sub-diffusive plateau of $\Delta r (t)$ is well established and spreads out over two decades.  Again, we observe a local maximum of $\chi_{\rm ss}(a, t)$ near the onset of the diffusive regime.  However, in addition to this ``cage-escaping" maximum we also observe a secondary ``cage-exploring" maximum, marked by an open circle, near the onset of the plateau; the dynamics are also heterogeneous on time scales when particles are becoming constrained by their neighbors.  Thus, in the regime where there are two distinct crossovers in the dynamics--one from ballistic to sub-diffusive motion, and one from sub-diffusive to diffusive motion--there are also two maxima in $\chi_{\rm ss}(a,t)$, located at distances $a$ and times $t$ corresponding to the two crossovers.

For the lowest temperature shown (fig.~\ref{maps}(c)), we do not reach a diffusive regime on the time scale of our simulation, which means that this temperature lies below the dynamic glass transition temperature $T_g\approx 5\times 10^{-4}$.  We therefore do not observe the cage-escaping maximum associated with the onset of the diffusive regime.  However, the location of the cage-exploring maximum remains relatively unchanged, as shown by the open circle.  Although the system is out of equilibrium for all $T<T_g$, we observe mild aging effects only for the highest of these temperatures studied, $T=4\times10^{-4}$ and $T=4.5\times10^{-4}$, where the equilibration times are comparable to the quench times, waiting times, and run times of our simulations.

\begin{figure}[t]
\onefigure[width=80mm]{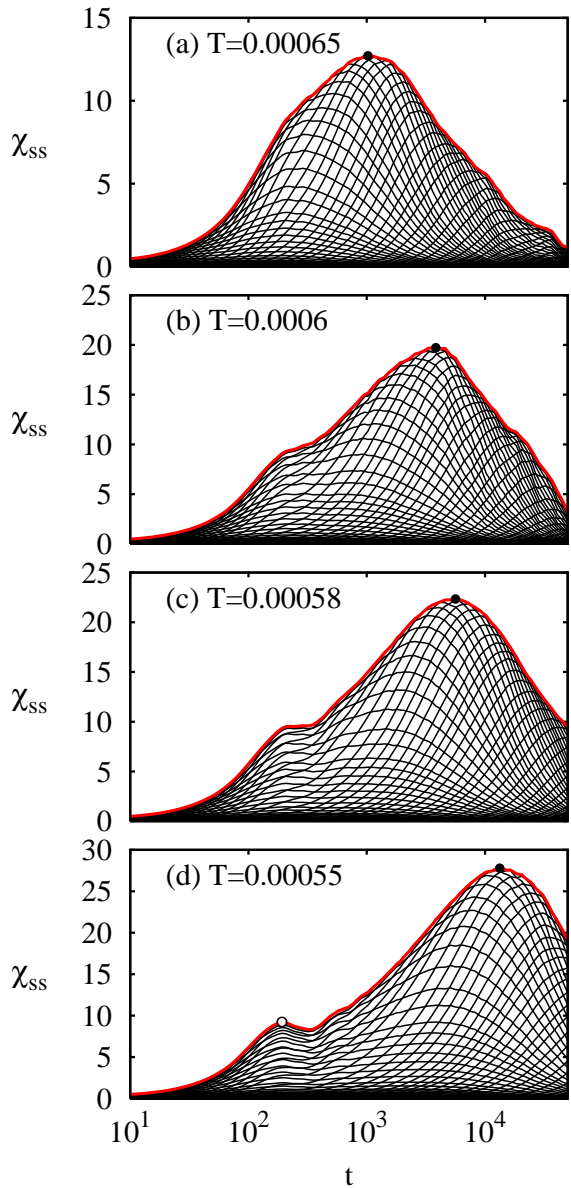}
\caption{Dynamic susceptibility $\chi_{\rm ss}(a, t)$ vs lag time $t$ for $p=10^{-2}$, $N=1600$, and four different temperatures above but near the glass transition: (a) $T=6.5\times 10^{-4}$, (b) $T=6\times 10^{-4}$, (c) $T=5.8\times 10^{-4}$, and (d) $T=5.5\times 10^{-4}$.  At each temperature, we show multiple curves, each corresponding to a different value of the overlap distance $a$.  The range of overlap distances represented by the curves is the same as in the color plots of fig.~\ref{maps}.  The maximum of $\chi_{\rm ss}$ with respect to $a$ at each value of $t$ is represented by the heavy red curve.  Circles mark the lag time $t^\star$ and amplitude $\chi^\star$ of local maxima of $\chi_{\rm ss}(a, t).$  As in fig.~\ref{maps}, filled circles represent cage-breaking local maxima near the end of the sub-diffusive plateau of $\Delta r(t)$, and open circles represent cage-forming local maxima near the beginning of the plateau.}
\label{plots}
\end{figure}

To study the evolution of the maxima with temperature, we plot in Fig.~\ref{plots} $\chi_{\rm ss}(a,t)$ as a function of $t$ for many different fixed values of $a$ covering the same range of $a$ as in Fig.~\ref{maps}.  The thick red curve marks the maximum of $\chi_{\rm ss}$ with respect to $a$ at each $t$.  Maxima of this curve (marked with circles) therefore represent maxima of $\chi_{\rm ss}$ with respect to both $a$ and $t$.  Fig.~\ref{plots} shows that at high temperatures, only one local maximum of $\chi_{\rm ss}$ can be resolved.  This is the well-known maximum at times $t$ comparable to the relaxation time; it is represented by a filled circle in fig.~\ref{plots}(a).  As $T$ is lowered, however, a shoulder becomes apparent at time scales corresponding to cage exploring, as shown in fig.~\ref{plots}(b).  This shoulder evolves into a secondary maximum, denoted by an open circle in fig.~\ref{plots}(c), as $T$ is lowered still further.  Note that the primary cage-escaping maximum is quite broad in $t$ and that the secondary cage-exploring maximum is only observed at temperatures at which the cage-escaping maximum has moved to sufficiently long times.  Thus, the reason why the observation of two distinct maxima is restricted to temperatures fairly close to the glass transition is because the smaller cage-exploring maximum is obscured by the larger cage-escaping maximum at higher temperatures.

\begin{figure}
\onefigure[width=80mm]{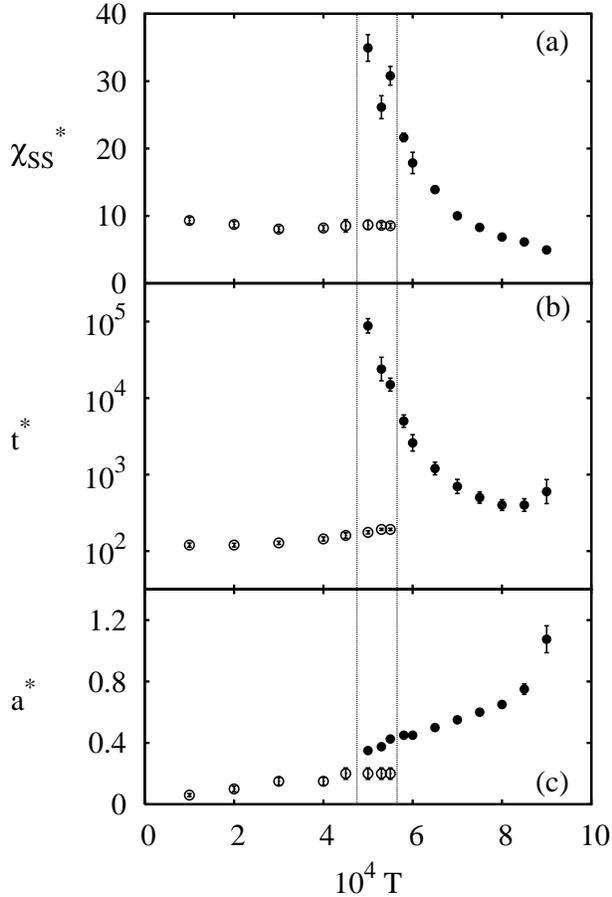}
\caption{Local maxima of the dynamic susceptibility.  We plot (a) the magnitudes of all observed local maxima of $\chi_{\rm ss}(a, t)$, $\chi^\star$, (b) their lag times $t^{\star}$, and (c) their overlap distances $a^\star$ as a function of temperature at fixed pressure $p=10^{-2}$ and system size $N=1600$.  As in fig.~\ref{maps}, the filled circles represent cage-escaping maxima, while the open circles represent cage-exploring maxima.  The vertical dashed lines denote the range over which we observe both types of maxima.  Note that (a) and (c) are presented on a linear scale, while (b) is presented on a log-linear scale.  Error bars represent the larger of (1) the standard deviation of the mean calculated from simulations with different initial conditions and (2) the uncertainty associated with the discrete sampling of $a$ and $t$.}
\label{max}
\end{figure}

Fig.~\ref{max} shows the temperature dependence of the two maxima of the dynamic susceptibility.  We plot the amplitude of the maxima $\chi^\star$ and their associated locations in time and space, $t^\star$, and $a^\star$, as functions of $T$ at fixed pressure $p=10^{-2}$ and system size $N=1600$.  The amplitude $\chi^\star$ and lag time $t^\star$ of the primary, cage-escaping maximum (solid symbols) both increase as $T$ decreases until $t^\star$ passes beyond our observable time window, in accord with many previous observations.  The overlap distance $a^\star$ of the cage-escaping maximum is on the order of half a disk diameter for temperatures near the glass transition, a distance about two times larger than the plateau of the rms displacement.  This distance presumably reflects the scale of the rearrangements necessary for diffusive motion to occur.   At high temperatures above $T=6 \times 10^{-4}$, the plateau in the rms displacement is less well defined and the cage-escaping maximum lies at slightly larger values of $a^\star$. 

The temperature dependence of the cage-exploring, secondary maximum is markedly different from that of the primary maximum.  Figure~\ref{max} shows that $\chi^\star$ and $t^\star$ remain roughly constant as temperature decreases from $T=5.5\times 10^{-4}$, just above $T_g \approx 5 \times 10{-4}$, to $T=10^{-4}$, which is well below it.  This is consistent with the behavior of the rms displacement; the time of the onset of the plateau does not vary much with temperature as long as the temperature is low enough for the plateau to exist.  The cage-exploring overlap distance $a^\star$ decreases with decreasing $T$, remaining near the plateau value in the rms displacement.

\begin{figure}
\onefigure[width=80mm]{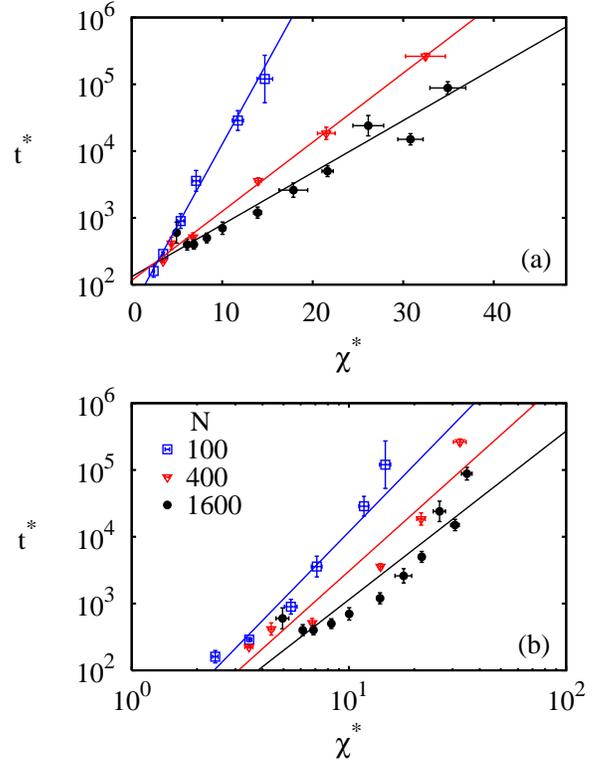}
\caption{Dependence of the relaxation time $t^\star$ on the spatial extent of kinetic heterogeneities, $\chi^\star$.  We plot the position of the cage-escaping maximum, $t^\star$ vs the magnitude of that maximum, $\chi^\star$ at fixed pressure $p=10^{-2}$.  Each point corresponds to a different temperature above $T_g$.  Results are shown for three different system sizes $N$, as labelled, on both (a) a log-linear scale and (b) a log-log scale.  The straight lines in (a) and (b) are exponential and power law fits, respectively.  Error bars represent the larger of (1) the standard deviation of the mean calculated from simulations with different initial conditions and (2) the uncertainty associated with the discrete sampling of $a$ and $t$.}
\label{tstar_chistar}
\end{figure}

The amplitude  $\chi^\star$ of the cage-escaping maximum is a measure of the number of disks whose dynamics are correlated at the crossover to diffusion, while $t^\star$ is a measure of the timescale for this longest, or $\alpha$-, relaxation process.  In fig.~\ref{tstar_chistar} we plot $\chi^\star$, the number of disks whose dynamics are correlated at the cage-escaping maximum of $\chi_{\rm ss}$, as a function of the time scale corresponding to the cage-escaping maximum, $t^\star$.  This time scale is a measure of the $\alpha$-relaxation process.   We have plotted $\chi^\star$ vs.~$t^\star$ for three different system sizes, $N=100$, 400, and 1600.  Note the substantial system-size dependence in $\chi^\star$~\cite{Karmakar2009}, even for temperatures for which the size of correlated regions $n^\star\approx 4 \chi^\star$ is only on the order of $10$.  This indicates that the finite size of the system affects the dynamics even when the relative linear size of the correlated regions is as small as $\sqrt{n^\star/N}\approx 0.1$.  This strong effect is likely due to a broad distribution of sizes of correlated regions.  We find that while the spatial correlation function of the overlap function has an average correlation length consistent with the value $\sqrt{\pi \langle \sigma^2\rangle n^\star/\phi}$ expected from a compact correlated region, it also has a long-ranged tail, decaying slower than exponentially with distance.  This long range tail reflects intermittent periods when the the clusters are extended.  We find that $t^\star$ and $a^\star$ also depend on system size.  If $a$ is held fixed, $t^\star$ decreases and $\chi^\star$ increases with increasing system size $N$ at all temperatures studied, consistent with Ref.~\cite{Karmakar2009}.  However, when we maximize over $a$, we find that at sufficiently high $T$, $t^\star$ increases with $N$ because $a^\star$ increases with $N$ there.

Fig.~\ref{tstar_chistar} shows that for each system size $N$, the dependence of $\chi^\star$ on $t^\star$ is qualitatively similar.  Note that the log-linear plot (fig.~\ref{tstar_chistar}(a)) is straighter than the log-log plot (fig.~\ref{tstar_chistar}(b)) for each system size, indicating that the dependence of $t^\star$ on $\chi^\star$ is described better by an exponential than by a power law.    Such an exponential dependence of relaxation time on cluster size is expected in scenarios involving cooperative rearrangements~\cite{Toninelli2005}.  Previous numerical studies of model glass formers fit $\chi^\star$ and $t^\star$ to power laws in $T-T_c$, where $T_c$ is a critical temperature~\cite{Glotzer2000, Lacevic2003b, Szamel2006}, implying that the relationship between $\chi^\star$ and $t^\star$ is also a power law.  However, these studies determined $\chi^\star$ as the maximum of $\chi_4(a, t)$ with respect to $t$ at fixed $a$, instead of searching for a maximum with respect to both $t$ and $a$.  Also, the relationship between $\chi^\star$ and $t^\star$ was not shown directly.  An analysis of a lower bound of $\chi_4$ for several real glass-forming liquids found an exponential dependence near the glass transition~\cite{Dalle-Ferrier2007}, consistent with our results.

\section{Discussion}

We quantified the size of kinetic heterogeneities for a model glass-forming liquid as a function of temperature in terms of fluctuations of a dynamic order parameter.  By measuring the dynamic susceptibility $\chi_{\rm ss}(a, t)$ as a function of both lag time and overlap distance, we identified two distinct local maxima.    The primary, ``cage-escaping" maximum, which has been found in previous numerical and experimental studies, occurs at the end of the plateau in the root-mean-square displacement, near the onset of diffusive transport.  We also discovered a secondary, ``cage-exploring'' maximum at the beginning of the plateau in the mean-squared displacement, which corresponds to the ballistic to sub-diffusive crossover.  Thus, the dynamics are most heterogeneous at displacements and times corresponding to crossovers in the dynamics.

The secondary, cage-exploring maximum at the crossover from ballistic to sub-diffusive motion emerges from the shoulder of the primary cage-escaping maximum when the temperature is low enough that there is a well-defined plateau in the mean-squared displacement.  We find that the plateau must extend over at least two orders of magnitude in order for the cage-exploring maximum to appear.  We observe both maxima over the range of temperatures that are low enough that the mean-squared displacement exhibits a broad plateau but high enough that the eventual crossover to the diffusive regime is still observable.  Note that although both maxima emerge, grow, and shift continuously with temperature, there is an apparent discontinuous drop in the largest observed maximum at the dynamical glass transition.  This occurs when the primary maximum passes out of observation range, leaving only the secondary, cage-exploring maximum (see fig.~\ref{max}(b)).  

While the crossover from sub-diffusive to diffusive motion measures the approach to the glass transition, the crossover from ballistic to sub-diffusive motion is a general and relatively innocuous feature of dense fluids and solids.  Our finding that both crossovers exhibit maxima in the dynamic susceptibility suggests that spatially heterogeneous dynamics are a general feature of dynamical crossovers caused by interactions.    Indeed, a crystalline system of monodisperse disks exhibits a cage-exploring maximum at the end of the ballistic regime that is similar to the one shown here.  We suspect that more exotic dynamical crossovers, like those between two diffusive regimes with different diffusion coefficients, should also exhibit kinetic heterogeneities.  The existence of kinetic heterogeneities is therefore a general feature of dynamical crossovers and should not necessarily be taken as a sign of an impending glass transition.  

\acknowledgments
We thank Adam Abate,  Olivier Dauchot, Douglas Durian, and Smarajit Karmakar for useful discussions.  This work was supported by DE-FG02-05ER46199 and in part by the MRSEC program under NSF-DMR05-20020.

\bibliographystyle{eplbib.bst}
%\bibliography{../../Bibliography/Bibliography}
\bibliography{manuscript.bbl}

\begin{thebibliography}{10}
\expandafter\ifx\csname url\endcsname\relax\def\url#1{\texttt{#1}}\fi

\bibitem{Glotzer2000b}
\Name{Glotzer S.~C.} \REVIEW{J. Non-Crystalline Solids }{274}{2000}{342}.

\bibitem{Richert2002}
\Name{Richert R.} \REVIEW{J. Phys. Condens. Matter }{14}{2002}{R703}.

\bibitem{Andersen2005}
\Name{Andersen H.~C.} \REVIEW{Proc. Natl. Acad. Sci. U.S.A. }{102}{2005}{6686}.

\bibitem{Berthier2005}
\Name{Berthier L., Biroli G., Bouchaud J.-P., Cipelletti L., Masri D.~E.,
  L'H\^{o}te D., Ladieu F. \and Pierno M.} \REVIEW{Science }{310}{2005}{1797}.

\bibitem{Dalle-Ferrier2007}
\Name{Dalle-Ferrier C., Thibierge C., Alba-Simionesco C., Berthier L., Biroli
  G., Bouchaud J.-P., Ladieu F., L'H\^{o}te D. \and Tarjus G.} \REVIEW{Phys.
  Rev. E }{76}{2007}{041510}.

\bibitem{Berthier2007a}
\Name{Berthier L., Biroli G., Bouchaud J.-P., Kob W., Miyazaki K. \and Reichman
  D.~R.} \REVIEW{J. Chem. Phys. }{126}{2007}{184503}.

\bibitem{Berthier2007b}
\Name{Berthier L., Biroli G., Bouchaud J.-P., Kob W., Miyazaki K. \and Reichman
  D.~R.} \REVIEW{J. Chem. Phys. }{126}{2007}{184504}.

\bibitem{Matsui1994}
\Name{Matsui J., Odagaki T. \and Hiwatari Y.} \REVIEW{Phys. Rev. Lett.
  }{73}{1995}{2452}.

\bibitem{Kob1995}
\Name{Kob W. \and Andersen H.~C.} \REVIEW{Phys. Rev. E }{52}{1995}{4134}.

\bibitem{Matsui1998}
\Name{Matsui J., Fujisaki M. \and Odagaki T.} \REVIEW{J. Non-Crystalline Solids
  }{235}{1998}{335}.

\bibitem{Hiwatari1998}
\Name{Hiwatari Y. \and Muranaka T.} \REVIEW{J. Non-Crystalline Solids
  }{235}{1998}{19}.

\bibitem{Glotzer2000}
\Name{Glotzer S.~C., Novikov V.~N. \and Schr$\o$der T.~B.} \REVIEW{J. Chem.
  Phys. }{112}{2000}{509}.

\bibitem{Lacevic2003b}
\Name{La\v{c}evi\'{c} N., Starr F.~W., Schr$\o$der T.~B. \and Glotzer S.~C.}
  \REVIEW{J. Chem. Phys. }{119}{2003}{7372}.

\bibitem{Toninelli2005}
\Name{Toninelli C., Wyart M., Berthier L., Biroli G. \and Bouchaud J.-P.}
  \REVIEW{Phys. Rev. E }{71}{2005}{041505}.

\bibitem{Szamel2006}
\Name{Szamel G. \and Flenner E.} \REVIEW{Phys. Rev. E }{74}{2006}{021507}.

\bibitem{Chandler2006}
\Name{Chandler D., Garrahan J.~P., Jack R.~L., Maibaum L. \and Pan A.~C.}
  \REVIEW{Phys. Rev. E }{74}{2006}{051501}.

\bibitem{Flenner2007}
\Name{Flenner E. \and Szamel G.} \REVIEW{J. Phys. Condens. Matter
  }{19}{2007}{205125}.

\bibitem{Charbonneau2007}
\Name{Charbonneau D. \and Reichman D.~R.} \REVIEW{Phys. Rev. Lett.
  }{99}{2007}{135701}.

\bibitem{Abate2007}
\Name{Abate A.~R. \and Durian D.~J.} \REVIEW{Phys. Rev. E }{76}{2007}{021306}.

\bibitem{Lechenault2008}
\Name{Lechenault F., Dauchot O., Biroli G. \and Bouchaud J.~P.}
  \REVIEW{Europhys. Lett. }{83}{2008}{46003}.

\bibitem{Evans1983a}
\Name{Evans D.~J. \and Morriss G.~P.} \REVIEW{Chem. Phys. }{77}{1983}{63}.

\bibitem{Karmakar2009}
\Name{Karmakar S., Dasgupta C. \and Sastry S.} \REVIEW{Proc. Natl. Acad. Sci.
  U.S.A. }{106}{2009}{3675}.

\end{thebibliography}

\end{document}